\documentclass[a4paper,11pt]{article} 
\input epsf.tex
\usepackage{graphics}

\newcommand{\be}{\begin{equation}}
\newcommand{\ee}{\end{equation}}
\newcommand{\ba}{\begin{eqnarray}}
\newcommand{\ea}{\end{eqnarray}}
\newcommand{\ban}{\begin{eqnarray*}}
\newcommand{\ean}{\end{eqnarray*}}

\newcommand{\braket}[2]{\mbox{$ \langle #1 | #2 \rangle $}}

\newcommand{\ket}[1]{\mbox{$ | #1 \rangle $}}
\newcommand{\bra}[1]{\mbox{$ \langle #1 | $}}

\newcommand{\si}{\sigma}

\newcommand{\one}{\leavevmode\hbox{\small1\normalsize\kern-.33em1}}

\newcommand{\moy}[1]{\langle #1 \rangle}

\begin{document}

\title{\vspace{6cm} \bf Optical telecom networks as weak quantum
measurements with post-selection}
\author{Nicolas Brunner\\ \\ Group of Applied Physics, University of Geneva \\ \\ Diploma thesis
\\ \\ Under the direction of N. Gisin and V. Scarani }

\date{\today}

 \maketitle

\newpage

\section{Introduction}

In this work we establish a link between two apparently unrelated
subjects: polarization effects in optical fibers and devices, and
the quantum theory of weak measurements \cite{weak}. We show that
the abstract concept of weak measurements followed by
post-selection, introduced a decade ago by quantum theorists,
naturally appears in the everyday physics of telecom networks.
\\
\\
Our analogy works as follows. First, polarization mode dispersion
(PMD) \cite{pmdpdl} performs polarization measurements by
spatially separating the fiber's eigenmodes. It turns out that the
usual telecom limit for PMD, where dispersion has to be minimized,
corresponds to the quantum regime of weak measurements. Then
polarization dependent losses (PDL) \cite{pmdpdl} perform
post-selection in a very natural way: one post-selects those
photons that have not been lost. This is non-trivial physics since
the losses depend precisely on the measured degree of freedom: the
polarization of light. In case of an infinite PDL (i.e. a
polarizer) the post-selection is done on a pure state. For a
finite PDL, the post-selection is done on a mixed state. Thus the
amount of PDL characterizes the kind of post-selection involved.
\\
\\
We show also that the quantum formalism of weak measurements can
simplify some "telecom" calculations and gives a better
understanding of the physics of networks. A telecom network can be
described as a concatenation of elements with PMD (fibers) and
some with PDL (couplers, isolators, etc). A simple formula for the
mean time-of-arrival for an arbitrary concatenation of PMD and PDL
elements is derived.

\section{Polarization mode dispersion}

Polarization mode dispersion (PMD) is the most important
polarization effect in optical fibers. It is due to the
birefringence of the fiber. In standard optics, a fiber is
represented by a channel supporting two polarization modes. The
main consequence of PMD, is that the time of flight along the
fiber will be different for each mode.
\\
\\
Before going into calculations we would like to clarify the
notations. We use the formalism of two dimensional Jones vectors
to describe polarization. In this representation a classical state
of polarization is equivalent to a quantum spin $\frac{1}{2}$. The
three typical pairs of polarizations - horizontal-vertical linear,
diagonal linear, left-right circular - are described respectively
by the eigenvectors of the Pauli matrices


\ba \si_{x} = \left( \begin{array}{cc}  0 & 1  \\ 1 & 0
\end{array} \right)  \quad \si_{y} = \left( \begin{array}{cc}  0 & -i  \\ i & 0
\end{array} \right) \quad \si_{z} = \left( \begin{array}{cc}  1 & 0  \\ 0 &
-1
\end{array} \right) \,.\ea

We mostly use the eigenstates of $\si_{z}$

 \ba \si_{z}\ket{H}=\ket{H} \quad \mbox{and} \quad
\si_{z}\ket{V}=-\ket{V} \ea

A general pure state of polarization is a complex superposition of
these two states. The state $\ket{+ \hat{n}}=\cos(\theta
/2)\ket{H} +\sin(\theta /2) e^{i \varphi}\ket{V}$ corresponds to
the point $\hat{n}=(\theta,\varphi)$ on the Poincar\'e sphere.
\\
\\
Let's consider a PMD fiber of length $L$ with birefringence vector
$\vec{\xi}$. We define the fiber's axis to be the $z$ direction.
So $\ket{H}$ and $\ket{V}$, the eigenstates of $\si_{z}$, are the
eigenmodes of the fiber. We consider a polarized gaussian pulse
with coherence time $t_{c}$. This pulse can be thought of as a
classical light pulse or as a quantum single photon
non-monochromatic state. Taking into account both energy and
polarization degrees of freedom, the complete initial state reads

\ba \ket{\Psi_{in}}\, = \frac{1}{(t_{c} \sqrt{2\pi})^{1/2}}\,e^{-
t^2 /4t_{c}^2 }\, \otimes\,\big(\alpha
\ket{H}+\beta\ket{V}\big)\,\equiv\, g(t)\otimes \ket{\psi_{0}}\,,
\ea

where $\alpha,\beta$ are complex numbers and satisfy $|\alpha|^2
+|\beta|^2=1$. In this equation, as usual, the gaussian function
$g(t)$ is normalized so that

\ba G(t)\,\equiv\, g(t)^2 = \frac{1}{t_{c} \sqrt{2\pi}}\,e^{-t^2
/2t_{c}^2} \ea

is a probability distribution. Our coordinate system $t$
corresponds to the time-of-arrival of the pulse and travels at the
speed of light in the fiber (without PMD) $v_{f}=c/n$, where $n$
is the refractive index in the fiber. A pulse centered at $t=0$
propagates at speed $v_{f}$.
\\
\\
PMD is a unitary operation represented by the operator

\ba U(\xi, \hat{z})=e^{i\frac{\xi}{2}\omega \si_{z}}=
\cos(\frac{\xi\omega}{2}) \one +i \sin(\frac{\xi\omega}{2})
\si_{z} \,\,. \ea

where $\xi=\,\parallel\vec{\xi}\parallel$. To compute the
evolution for each eigenmode of the fiber, we have to
Fourier-transform the input state $g(t)\otimes \ket{H,V}$ into the
frequency domain, apply the PMD operator to any monochromatic
component, and integrate back to the time domain. Note that by
$\ket{H,V}$, we mean $\ket{H}$ or $\ket{V}$. Thus we have

\ba U(\xi, \hat{z}) \,g(t)\otimes \ket{H,V} =\int d \omega
\tilde{g}(\omega) e^{-i \omega t}  \, e^{i\frac{\xi }{2}\omega
\si_z }\, \ket{H,V} \ea

where $\tilde{g}(\omega)$ is the Fourier-transform of $g(t)$.
Since our gaussian pulse is centered in frequency $\omega_{0}$ we
apply the following change of coordinate $\omega \rightarrow
\omega-\omega_{0}$. We have

\ba U(\xi, \hat{z}) \,g(t)\otimes \ket{H,V} &=& \int d \omega
\tilde{g}(\omega-\omega_{0}) e^{-i (\omega-\omega_{0}) t}  \,
e^{i\frac{\xi }{2}\omega  \si_z }\, \ket{H,V}  \nonumber \\
&=&   \int d \omega \tilde{g}(\omega-\omega_{0}) e^{-i
(\omega-\omega_{0}) (t \mp \frac{\xi }{2} )}  \, e^{\pm
i\frac{\xi}{2}\omega_{0} }\, \ket{H,V} \nonumber
\\
&=& g(t \mp \frac{\xi}{2} ) \,\otimes \, e^{\pm i\frac{\xi
}{2}\omega_{0} }\, \ket{H,V} \,\,.\ea

 The above equation shows
that the mean time of arrival for each eigenmode is shifted by the
same quantity ($\xi/2$) in either direction. Usually the time
separating the mean time-of-arrival of the two modes is called the
differential group delay (DGD) and noted $\delta \tau$ . Thus the
output state reads

\ba\label{psipmd} \ket{\Psi_{PMD}} =
 \tilde{\alpha} g_{-}(t)  \ket{H} + \tilde{\beta} g_{+}(t)  \ket{V} \,, \ea

 where we have omitted the tensor products and defined

\ba \tilde{\alpha} \equiv e^{i\frac{\delta\tau }{2}\omega_{0}
}\alpha \quad ,  \quad  \tilde{\beta} \equiv e^{-i\frac{\delta\tau
}{2}\omega_{0} }\beta \quad ,  \quad g_{\pm}(t) \equiv g(t \pm
\frac{\delta \tau }{2}) \,\,.
 \ea

 So the effect of PMD is a spatial separation of the fiber's
 eigenmodes $\ket{H}$ and $\ket{V}$ combined with a rotation
 of the polarization around the $z$ axis. This global phase is
 irrelevant for the link we want to establish and will therefore
 always be absorbed in the polarization state

 \ba \label{out} \ket{\psi} \equiv
 e^{i\omega_{0}\frac{\delta\tau}{2}\si_z}\ket{\psi_{0}}=
 \tilde{\alpha}\ket{H} + \tilde{\beta}\ket{V}\,\,.\ea

\begin{figure}[h!]
\centerline{\scalebox{.6}{\includegraphics{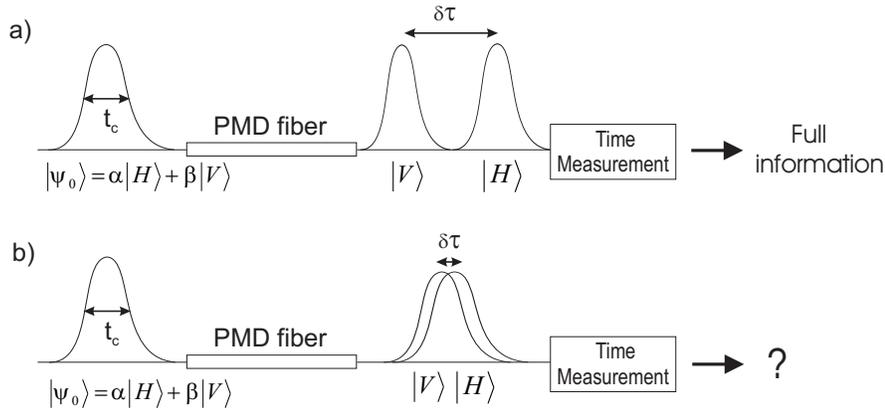}}}
\caption{PMD as a polarization measuring device. a) the
measurement is strong: the time-of-arrival gives full information
about the polarization of the pulse. b) the measurement is weak:
no discrimination possible.} \label{strongweak}
\end{figure}

 Note that $\ket{H}$ and
$\ket{V}$ are respectively the fastest and slowest polarization
modes in the fiber.
\\
\\
 The measurement analogy now goes like this.
 Different polarization modes will arrive at different times.
 Thus by measuring the mean time-of-arrival of a pulse one can obtain
 some information about its state of polarization. Depending on whether
 the delay between the two polarization modes is large
 (small) compared to the coherence time of the pulse,
 the measurement of polarization achieved by PMD is strong
(weak) (see Fig. \ref{strongweak}). In telecom optics the
interesting limit is always
 the second one since dispersion has to be minimized.

In standard quantum mechanics, a measurement is an interaction
between two systems: the physical system to be measured and the
measuring device, usually called a pointer \cite{peres}. The
change in the position of the pointer, due to the measurement,
provides some information about the state of the measured system.
In our case the energy degree of freedom of the pulse plays the
role of the pointer, while the polarization is the measured
system. Thus pointer and measured system are here different
degrees of freedom of the {\em same} physical system. PMD provides
the necessary interaction between the pointer and the measured
system since energy and polarization degrees of freedom are
entangled in the output state (\ref{psipmd}). However, PMD alone,
without the final measurement of the time-of-arrival, is not a
measurement since it is a reversible operation.

\newpage


\section{Polarizer: post-selection on a pure state }



\subsection{Strong measurement}

Our aim is to describe networks. A network is an arbitrary
concatenation of PMD and PDL elements. In this section we study
the simplest one: a PMD fiber followed by an element with infinite
PDL, i.e. a polarizer. Note that this setup corresponds to the
usual scenario in which weak measurements are studied, since the
system is pre- and post-selected. Here the post-selection is done
on a pure state, since the polarizer projects onto a linear
polarization

\begin{displaymath}
\ket{\phi} = \mu \ket{H} + \nu \ket{V}\,.
\end{displaymath}

At the output, we measure the time-of-arrival. So, the final state
is

\ba\label{psiout} \ket{\Psi_{out}} =
\ket{\phi}\braket{\phi}{\Psi_{PMD}}= \Big[\tilde{\alpha} \bar{\mu}
\, g_{-}(t) +\tilde{\beta} \bar{\nu} \,g_{+}(t)
\Big]\otimes\ket{\phi}\,\equiv\, F(t)\otimes \ket{\phi}\,. \ea

Note that $|F(t)|^2$ must be normalized in order to be a
probability distribution. As we have seen above, because of PMD
some information about the preparation is contained in the shape,
$F(t)$, of the output pulse.
\\
\\
When $\delta \tau>>t_{c}$ the two gaussians do not overlap,
$g_{-}(t) g_{+}(t) \approx 0$, and it is possible to discriminate
between the two polarization eigenmodes. The detected intensity
corresponds to two well-separated gaussians. From (\ref{psiout})
we have

\ba I(t)\,\equiv\,|F(t)|^2&=&
|\tilde{\alpha}\bar{\mu}|^2\,G_{-}(t)
\,+\,|\tilde{\beta}\bar{\nu}|^2\,G_{+}(t)\,=\nonumber\\
&=& \mbox{Prob}(\phi|H)\mbox{Prob}(H|\psi)\,G_{-}(t)
\,+\,\mbox{Prob}(\phi|V)\mbox{Prob}(V|\psi)\,G_{-}(t) \ea

where we have introduced obvious notations like
$\mbox{Prob}(H|\psi)=|\tilde{\alpha}|^2$ etc. The probability that
the polarization was $\ket{H}$, given the preparation and the
post-selection, is, in the limit $\delta \tau>>t_{c}$

\ba
\mbox{Prob}(H)&=&\frac{\int_{0}^{\infty}I(z)dz}{\int_{-\infty}^{\infty}I(z)dz}\,=\,
\frac{\mbox{Prob}(\phi|H)\mbox{Prob}(H|\psi)}{\sum_{K=H,V}
\mbox{Prob}(\phi|K)\mbox{Prob}(K|\psi)}\,. \ea

This is exactly the Aharonov-Bergmann-Lebowitz (ABL) rule, which
is in fact the classical rule for the probability of sequential
events. Of course, $\mbox{Prob}(V)=1-\mbox{Prob}(H)$. One can also
compute the mean time of arrival

\begin{equation}
\langle t \rangle = \mbox{Prob}(H)\frac{\delta \tau}{2}
+\mbox{Prob}(V)\big(-\frac{\delta \tau}{2} \big)\,.
\end{equation}

Since the mean value of $\si_{z}$ is simply $\langle \sigma_{z}
\rangle=\mbox{Prob}(H)-\mbox{Prob}(V)$, we have

\begin{equation}\label{main}
 \langle t \rangle \,=\, \frac{\delta \tau}{2}
\, \langle \sigma_{z} \rangle\,.
\end{equation}

The interpretation of equation (\ref{main}) is a key point of our
work and needs to be carefully explained. In quantum-mechanical
terms, $\sigma_z$ is the observable that is measured by the time
delay $\delta\tau$, and (\ref{main}) shows that its mean value is
associated to the mean time-of-arrival $\langle t \rangle$ in the
case of a strong measurement. What happens now when the
measurement is weakened? The point is that, in contrast to
$\mbox{Prob}(H)$ and $\mbox{Prob}(V)$, the mean time-of-arrival is
a physical quantity that can be defined and measured in any
situation (for a strong or weak measurement, with or without
post-selection). We admit that (\ref{main}) is {\em the definition
of the mean value of $\sigma_z$ when measured by introducing a
time delay $\delta\tau$ between $\ket{H}$ and $\ket{V}$.}

\subsection{General measurement}

We can now remove all assumptions on the {\em strength} of the
measurement and derive an analytical formula for the mean
time-of-arrival $\langle t \rangle$. Lets consider again the
output state (\ref{psiout}). Without any assumption on the
gaussians, the intensity is now

\ba I(t)\,\equiv\,|F(t)|^2 = |A|^2\,G_{-}(t) \,+\,|B|^2\,G_{+}(t)
+ 2 Re(\bar{A} B) g_{-}(t) \,g_{+}(t) \,, \ea

where $A\equiv\tilde{\alpha} \bar{\mu}$ and $B\equiv \tilde{\beta}
\bar{\nu}$. The mean time-of-arrival is given by

\ba \langle t \rangle = \frac{\int_{-\infty}^{\infty}t  I(t)dt}
{\int_{-\infty}^{\infty}I(t)dt} = \frac{(|A|^2-|B|^2) \frac{
\delta \tau }{2}} {|A|^2+|B|^2+ 2 Re(\bar{A} B)
\int_{-\infty}^{\infty} g_{-}(t) \,g_{+}(t) dt} \,. \ea

We evaluate separately the remaining integral

\ba \label{int} \int_{-\infty}^{\infty} g_{-}(t) \,
  g_{+}(t)\,  dt = e^{-\frac{1}{2}({\delta \tau
}/{2t_{c}})^2} \underbrace{\frac{1}{t_{c} \sqrt{2 \pi}}
\int_{-\infty}^{\infty} e^{-{ t^2}/{2t_{c}^2}} \, dt}_{=1}  =
e^{-\frac{1}{2}({\delta \tau }/{2t_{c}})^2}\,. \ea

So finally we find

\ba\label{tofgen} \langle t \rangle= \frac{ \delta \tau }{2} \,
\frac{(|A|^2-|B|^2)} {|A|^2+|B|^2+ 2 Re(\bar{A} B) \,
e^{-\frac{1}{2}({\delta \tau }/{2t_{c}})^2}} \,. \ea

An important feature of equation (\ref{tofgen}) is that the
dependance in the {\em strength} of the measurement (i.e. in
$\delta \tau /t_{c}$) is very explicit. Of course in the limit
$\delta \tau /t_{c} \rightarrow \infty $, corresponding to a
strong measurement our previous result is recovered.
\\
\\
Since equation (\ref{tofgen}) is completely general we can compute
the mean time-of-arrival in the case of a weak measurement,
corresponding to the telecom limit of PMD. When
$\delta\tau<<t_{c}$ equation (\ref{tofgen}) becomes

\ba\ \langle t \rangle= \frac{ \delta \tau }{2} \,
\frac{(|A|^2-|B|^2)} {|A|^2+|B|^2+ 2 Re(\overline{A} B)} = \frac{
\delta \tau }{2} \, Re \big( \frac{A-B} {A+B}\big) \ea

Note that

\ba A \pm B= \tilde{\alpha} \bar{\mu} \pm \tilde{\beta} \bar{\nu}
=
\left\{ \begin{array}{l}  \braket{\phi}{\psi}  \\
\bra{\phi} \si_{z} \ket{\psi}\end{array} \right. \,.\ea

Using (\ref{main}), we find

\ba \label{weakpure} \moy{\si_z}_{w}= \, Re \Big( \frac{\bra{\phi}
\si_{z} \ket{\psi}} {\braket{\phi}{\psi}} \Big) \,, \ea

which is exactly the weak value of $\si_{z}$ when the
post-selection is done on a pure state $\ket{\phi}$, according to
Aharonov and Vaidman \cite{weak}. Note that $\moy{\si_z}_{w}$ can
reach arbitrarily large values, leading to an apparently
paradoxical situation since the eigenvalues of $\si_z$ are $\pm1$.
But there is no paradox at all since $\moy{\si_z}_{w}>1$ simply
means $\moy{t}>\frac{\delta\tau}{2}$. This situation is reached by
post-selecting on a state $\ket{\phi}$ nearly orthogonal to
$\ket{\psi}$. These are very rare events; the shape $F(t)$ of the
pulse is strongly distorted, and it is not astonishing that its
{\em center of mass} could be found far away from its expected
position in the absence of post-selection.

\section{PDL: post-selection on a mixed state }

\subsection{General measurement}


We now go one step further into the description of a general
network and replace the polarizer of the previous section by an
element with finite PDL, for example a coupler, an isolator, an
amplifier or a circulator. Neglecting a global attenuation, PDL is
represented by a non-unitary operator

\ba F(\mu, \hat{n})=e^{\mu \si_{n}/2}=\cosh(\frac{\mu}{2}) \one +
\sinh(\frac{\mu}{2}) \si_{n} \ea

where $\si_{n}=\hat{n} \cdot \vec{\si}$. The most and least
attenuated states, respectively $\ket{-\hat{n}}$ and
$\ket{+\hat{n}}$, are orthogonal. The attenuation between them,
expressed in dB, is $10 \log_{10}(e^{2\mu})$. Mathematically
speaking the PDL operator is not a projective measurement, but a
more general operation called a POVM \cite{peres}. In quantum
theory, it is usually called a filter, for example in the
unambiguous discrimination of non-orthogonal quantum states
\cite{unambig}.
\\
\\
As in the previous section, we derive now a formula for the mean
time-of-arrival. To simplify the notation we define $F\equiv
F(\mu, \hat{n})$. The output state is now

\ba\label{psiout2} \ket{\Psi_{out}} = F \ket{\Psi_{PMD}}=
A(t)\ket{H}+ B(t)\ket{V}\,, \ea

where

\ba A(t)&=& \bra{H}F\ket{H}\tilde{\alpha}g_{-}(t)+
\bra{H}F\ket{V}\tilde{\beta}g_{+}(t) =
(C+n_{z}S)\tilde{\alpha}g_{-}(t) + Sn_{-}\tilde{\beta}g_{+}(t)
\\ B(t)&=& \bra{V}F\ket{V}\tilde{\beta}g_{+}(t)+
\bra{V}F\ket{H}\tilde{\alpha}g_{-}(t) =
(C-n_{z}S)\tilde{\beta}g_{+}(t) + Sn_{+}\tilde{\alpha}g_{-}(t) \ea

with $C \equiv\cosh(\frac{\mu}{2})$, $S\equiv\sinh(\frac{\mu}{2})$
and $n_{\pm}=n_{x}\pm in_{y}$. We compute the output intensity

\ba I(t)= |A(t)|^2+ |B(t)|^2 &=& |\tilde{\alpha}|^2 (\cosh\mu+
n_{z}\sinh(\mu))G_{-}(t)+ |\tilde{\beta}|^2 (\cosh\mu-
n_{z}\sinh\mu)G_{+}(t) \nonumber\\ & & + \,\, 2 \sinh \mu
Re(\tilde{\alpha}\bar{\tilde{\beta}}n_{+}
e^{i\delta\tau\omega_{0}})g_{+}(t)g_{-}(t)\,\, . \ea

Using again (\ref{int}) the mean time-of-arrival is

\ba \langle t \rangle = \frac{\int_{-\infty}^{\infty}t  I(t)dt}
{\int_{-\infty}^{\infty}I(t)dt} = \frac{ \delta \tau }{2}
\frac{|\alpha|^2-|\beta|^2+ \gamma n_{z}}  {1+ \gamma  \, \big[
n_{z} ( |\alpha|^2-|\beta|^2)+  2
Re(\alpha\bar{\beta}n_{+}e^{i\delta\tau\omega_{0}})
e^{-\frac{1}{2}({\delta \tau }/{2t_{c}})^2} \big] } \,, \ea

where we have defined $\gamma  \equiv \tanh \mu$. In the limit of
a weak measurement and using equation (\ref{main}) we find

\ba \label{weak1} \moy{\si_z}_{w}= \frac{\moy{\si_z}_{\psi }+
\gamma n_{z}} {1+ \gamma \hat{n} \cdot\vec{ \si}} =\, Re \Big(
\frac{\bra{\psi} F^{\dag}F \si_{z} \ket{\psi}}
{\bra{\psi}F^{\dag}F \ket{\psi}} \Big) =\, Re \Big(
\frac{\bra{\psi} F^{2} \si_{z} \ket{\psi}} {\bra{\psi}F^2
\ket{\psi}} \Big)\,, \ea

where we have used the fact that $F$ is self-adjoint, i.e.
$F^{\dag}=F$. This is exactly the expression given by quantum
theorists for the mean value of $\si_{z}$ when post-selection is
done on the mixed state $\rho= \frac{1}{Tr(F^2 )}F^2 $. Note
however that when a photon comes out, it is left in the pure state
$F\ket{\psi}$. So the meaning of {\em post-selection on a mixed
state} has to be explained. In the theory of weak measurements,
the state of the system at the time of the intermediate weak
measurement is determined by two different pieces of information:
one coming from the past, i.e. the state in which the system was
pre-selected, and one coming from the future, the post-selected
state. In our case this second piece of information is a mixed
state since it corresponds to having the identity evolve back
through the system, i.e. the mixed state $\rho$.
\\
\\
It is interesting to work out the limiting cases of equation
(\ref{weak1}). When there is no PDL at all, $\gamma=0$ and we
recover $\moy{\si_z}_{w} =\moy{\si_z}_{\psi}$ since there is no
post-selection. For $\gamma=1$ which corresponds to an infinite
PDL $(\mu \rightarrow \infty )$ we recover our previous result for
the post-selection on a pure state (\ref{weakpure}), with
$\ket{\phi}=\ket{+\hat{n}}$.

\subsection{Anomalous dispersion and principal states of polarization}

As $\moy{\si_z}_{w}$ can be larger than one we recover anomalous
dispersion, which was one of the major results of combining the
effects of PMD and PDL \cite{pmdpdl}. From (\ref{weak1}) the
maximum (and minimum) mean time-of-arrival can be computed. The
largest value is obtained when PDL is orthogonal to PMD, say
$\hat{n}=\hat{x}$. Varying over the input polarization
$\ket{\psi}=\alpha\ket{H }+\beta \ket{V}$, we find

\ba \label{max} \mbox{max/min}_{ \{ \alpha,\beta \} }\, \langle t
\rangle = \pm \frac{ \delta \tau }{2} \frac{1}{\sqrt{1-\gamma^2}}
\,, \ea

which corresponds to the results of \cite{pmdpdl}. Note however
that the factor $\gamma$ appearing in this last equation has a
completely different meaning in \cite{pmdpdl}. It is defined as
the overlap of the principal states of polarization (PSP), which
are the polarization states such that the output polarization is
independent of the optical frequency in first order. The concept
of PSP plays a key role in the usual PMD-PDL theory. It is quite
interesting to see that we recover these states in the quantum
approach. In fact they are simply the PMD fiber's eigenmodes after
their evolution through the setup, i.e. $F\ket{H}$ and $F\ket{V}$.
One can easily check that their overlap $\bra{V}F^2 \ket{H}$ is
equal to $\gamma$ and that their mean time-of-arrival is $\pm
\delta\tau/2$.

\section{Quantum formalism for optical networks}

\subsection{General network}

\begin{figure}[h]
\centerline{\scalebox{.4}{\includegraphics{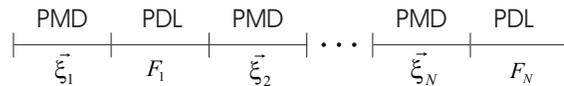}}}
\caption{General optical network. A concatenation of PMD and PDL
elements.} \label{concat}
\end{figure}

In this section we show that our work is more than a beautiful
analogy and that the quantum formalism can greatly simplify some
telecom calculations. We compute the mean time-of-arrival for an
arbitrary concatenation of PMD and PDL elements (see
Fig.\ref{concat}). Each PMD section is characterized by a
birefringence vector $\vec{\xi}_{n}$, its norm being equal to the
DGD, $\delta\tau_{n}$, and its direction specifying the measured
observable $\frac{1}{\delta\tau_{n}} \, \vec{\xi}_{n} \cdot
\vec{\si}\equiv \si_{n}$. Each PDL element is represented by an
operator $F_{n}$. Our input state is again a polarized gaussian
pulse $\ket{\Psi_{in}}=\ket{g(t)}\otimes \ket{\psi}$ of central
frequency $\omega_{0}$, with coherence time $t_{c}$. Considering
energy and polarization degrees of freedom the output state is

 \ba \ket{\Psi_{out}} = F_{N}\, e^{i\omega \frac{\delta\tau_{N}}{2}\si_{N}}
\cdots F_{1}\, e^{i\omega \frac{\delta\tau_{1}}{2}\si_{1}}  \,\,
\ket{g(t)} \otimes \ket{\psi} = \prod_{j=N}^{1} F_{j}\, e^{i\omega
\frac{\delta\tau_{j}}{2}\si_{j}}\, \ket{g(t)} \otimes \ket{\psi}
\,\,. \ea

The product operation has to be carefully defined since operators
$F_{j}$ do not commute. We use the notation

\ba \prod_{j=1}^{N} F_{j}\equiv F_{1}F_{2}\cdots F_{N} \quad
\mbox{and} \quad  \prod_{j=N}^{1} F_{j}\equiv F_{N}F_{N-1}\cdots
F_{1} \,\,.\ea

As it was shown in section 2, the effect of each PMD section is a
spatial separation of the fiber's eigenmodes and a global rotation
around the fiber's axis. Since both operations commute we can
rewrite

\ba  \ket{\Psi_{out}} = \prod_{j=N}^{1} F_{j}\, e^{i\omega
\frac{\delta\tau_{j}}{2}\si_{j}}\, \ket{g(t)} \otimes \ket{\psi} =
\prod_{j=N}^{1} \tilde{F}_{j} \, e^{i \delta\omega
\frac{\delta\tau_{j}}{2}\si_{j}}\, \ket{g(t)} \otimes
\ket{\psi}\,\,, \ea

where we have defined $\tilde{F}_{j} \equiv F_{j} \, e^{i
\omega_{0} \frac{\delta\tau_{j}}{2}\si_{j}}$. Note that
$\tilde{F}$ is not self-adjoint. Since we work in the telecom
limit of PMD, which we have shown to be equivalent to the regime
of weak measurements, all DGD's $\delta\tau_{n}$ are assumed to be
small compared to the coherence time $t_{c}$ of the optical pulse.
Therefore the exponential in the above equation can be expanded
since $\delta\omega \delta\tau_{j}= \delta\tau_{j}/t_{c}<<1$. We
have

\ba  \label{frete} \ket{\Psi_{out}} &\simeq& \prod_{j=N}^{1}
\tilde{F}_{j} \, (\one +i \delta\omega
\frac{\delta\tau_{j}}{2}\si_{j})\, \, \ket{g(t)} \otimes
\ket{\psi} \nonumber \\ &=& \bigg[(\prod_{j=N}^{1} \tilde{F}_{j})
+ i \delta\omega \sum_{j=1}^{N} \frac{\delta\tau_{j}}{2}
F_{N}F_{N-1} \cdots F_{j}\si_{j}F_{j-1} \cdots F_{1} \bigg] \, \,
\ket{g(t)} \otimes \ket{\psi} \,\,. \ea

We now write the state in a well-chosen polarization basis, namely
$\big\{ \prod_{j=N}^{1} \tilde{F}_{j}\ket{\psi}\,\, , \,\,
\ket{\mbox{orthog}} \big\} $, where $\ket{\mbox{orthog}}$ is
defined by $\bra{\mbox{orthog}} \prod_{j=N}^{1}
\tilde{F}_{j}\ket{\psi}=0 $ . Note that since $F_{j}$ are
non-unitary operations, the state $\prod_{j=N}^{1}
\tilde{F}_{j}\ket{\psi}$ is not normalized. We write its norm $B
\equiv \bra{\psi}\prod_{j=1}^{N} \tilde{F}_{j}^{\dag} \,
\prod_{j=N}^{1} \tilde{F}_{j}\ket{\psi}$. In equation
(\ref{frete}) we insert the completeness relation

\ba \one= \frac{1}{B^2}\prod_{j=N}^{1} \tilde{F}_{j}\ket{\psi}
\bra{\psi} \prod_{j=1}^{N} \tilde{F}_{j}^{\dag} +
\ket{\mbox{orthog}} \bra{\mbox{orthog}}\,\,, \ea

so that

\ba  \ket{\Psi_{out}} &=& \frac{1}{B^2}\prod_{j=N}^{1}
\tilde{F}_{j}\ket{\psi} \, \bigg[ \bra{\psi}\prod_{j=1}^{N}
\tilde{F}_{j}^{\dag} \,\bigg( (\prod_{j=N}^{1} \tilde{F}_{j}) + i
\delta\omega \sum_{j=1}^{N} \frac{\delta\tau_{j}}{2} F_{N} \cdots
F_{j}\si_{j}F_{j-1} \cdots
F_{1} \bigg) \bigg]  \ket{g(t)} \otimes \ket{\psi}  \nonumber\\
&+& \ket{\mbox{orthog}} \,\, \bigg[ \, \bra{\mbox{orthog}}  \,
\bigg( (\prod_{j=N}^{1} \tilde{F}_{j}) + i \delta\omega
\sum_{j=1}^{N} \frac{\delta\tau_{j}}{2} F_{N} \cdots
F_{j}\si_{j}F_{j-1} \cdots F_{1} \bigg) \bigg]  \ket{g(t)} \otimes
\ket{\psi}\,\,. \ea

The probability of finding the photon in the state
$\ket{\mbox{orthog}}$ is of order
$O(\delta\omega\delta\tau_{j})^2$, and therefore negligible. So
whenever the photon manages to come out it is left in the pure
state

\ba  \label{strf} \ket{\Psi_{out}} &=&  \bigg[ \, \one
+i\delta\omega W \, \bigg] \, \bigg( \ket{g(t)}
\otimes ( \prod_{j=N}^{1} \tilde{F}_{j}\ket{\psi}) \bigg) \nonumber\\
&\simeq& e^{i\delta\omega W } \, \bigg( \ket{g(t)} \otimes (
\prod_{j=N}^{1}
\tilde{F}_{j}\ket{\psi})  \bigg) \nonumber \\
&=& \underbrace{\bigg( B \, e^{-\delta\omega Im(W)
}\bigg)}_{\mbox{attenuation}} \,\,\,  \underbrace{ \bigg(
\ket{g(t+Re(W))} \otimes ( \frac{1}{B} \prod_{j=N}^{1}
\tilde{F}_{j}\ket{\psi}) \bigg)}_{\mbox{normalized state }}\,\,,
\ea

where

\ba \label{W} W = \sum_{j=1}^{N} \frac{\delta\tau_{j}}{2}
 \,   \frac{\bra{\psi}\prod_{k=1}^{N} \tilde{F}_{k}^{\dag}\,
     \, \prod_{k=N}^{j+1}
    \tilde{F}_{k} \, \si_{j} \prod_{k=j}^{1}
\tilde{F}_{k}\ket{\psi} } {\bra{\psi}\prod_{j=1}^{N}
\tilde{F}_{j}^{\dag} \, \prod_{j=N}^{1} \tilde{F}_{j} \,
\ket{\psi} \, } \ea

So the mean time-of-arrival is

\ba \label{tof5} \langle t \rangle = \sum_{j=1}^{N}
\frac{\delta\tau_{j}}{2}
 \,  Re \bigg( \frac{\bra{\psi}\prod_{k=1}^{N} \tilde{F}_{k}^{\dag}\,
     \, \prod_{k=N}^{j+1}
    \tilde{F}_{k} \, \si_{j} \prod_{k=j}^{1}
\tilde{F}_{k}\ket{\psi} } {\bra{\psi}\prod_{j=1}^{N}
\tilde{F}_{j}^{\dag} \, \prod_{j=N}^{1} \tilde{F}_{j} \,
\ket{\psi} \, }\bigg) \,\,. \ea

Note that the state (\ref{strf}) is not normalized because of the
losses in the system, and that the attenuation depends on the
central frequency of the pulse $\omega_{0}$. Expression
(\ref{tof5}) may seem quite cumbersome but can be rewritten in a
far more intuitive way. We define the following notation

\ba w(\vec{\xi_{j}},\ket{\psi},\rho) \equiv
\frac{\delta\tau_{j}}{2} Re \bigg(\frac{\bra{\psi} \rho \, \si_{j}
\ket{\psi}} {\bra{\psi} \rho \ket{\psi}} \bigg)\,\,. \ea

 So $w(\vec{\xi},\ket{\psi},\rho)$ is the mean time shift of a
PMD section with birefringence vector $\vec{\xi}$ when the input
state is $\ket{\psi}$ and the post-selection is done on the mixed
state $\rho$. With this, equation (\ref{tof5}) reads

\ba \label{general} \langle t \rangle &=& w
\big(\vec{\xi}_{1}\,,\,\ket{\psi}\,,\,\tilde{F}_{1}^{\dag} \cdots
\tilde{F}_{N}^{\dag} \tilde{F}_{N} \cdots \tilde{F}_{1}\big) \,+\,
w \big( \vec{\xi}_{2}\,,\, \tilde{F}_{1}\ket{\psi}\,,\,
\tilde{F}_{2}^{\dag} \cdots \tilde{F}_{N}^{\dag} \tilde{F}_{N}
\cdots \tilde{F}_{2}\big) \nonumber \\& &  + \cdots + \, w \big(
\vec{\xi}_{N}\,,\, \tilde{F}_{N-1} \cdots \tilde{F}_{1}
\ket{\psi}\,, \, \tilde{F}_{N}^{\dag} \tilde{F}_{N}\big) \ea

So the mean time-of-arrival is simply the sum of the contributions
of each PMD section computed when forgetting about all others
PMD's. This is quite natural since all PMD's are assumed to be
weak measurements, which means they modify only slightly the shape
of the pulse ($\delta\tau_{j}<<t_{c}$ $\forall j$).
\\
\\
So we obtain an analytical formula for the mean time-of-arrival
for an arbitrary concatenation of PMD and PDL elements. It should
be stressed that the structure of this formula is very simple. In
this sense we feel our analogy simplifies telecom calculation
since the equivalent computation in the usual PMD-PDL language is
less straightforward.
\\
\\
Another important result of \cite{pmdpdl} was that any
concatenation of PMD and PDL elements is equivalent to a simple
setup where an effective PMD is followed by an effective
frequency-dependent PDL. This is a consequence of the polar
decomposition theorem for complex matrices, which states that any
complex matrix $T$ can be decomposed into a unitary matrix $U$ and
a positive Hermitian one $A$, so that $T=AU$. Unfortunately we
were unable to recover this result in the quantum formalism.

\subsection{Optimal concatenation problem}

To illustrate the transparency of the structure of equation
(\ref{general}), we discuss the following problem. How should one
assemble a given number of PMD and PDL sections in order to
maximize (or minimize) the mean time-of-arrival $\langle t
\rangle$? In other words which setup optimizes the interaction
between PMD and PDL. For definiteness, we consider a five elements
PMD-PDL-PMD-PDL-PMD network. We omit all rotations due to PMD
sections, because they clearly play no role in our problem. Note
that anyway, one could change the lengths of each PMD element so
that the rotation due to PMD is a multiple of $2\pi$. Using
equation (\ref{general}) we have

\ba \label{num} \langle t \rangle =  \frac{\delta\tau_{1}}{2}
 \, \frac{  \bra{\psi}  F_{1} F_{2}^2 F_{1}\, \si_{1} \ket{\psi} }
    {\bra{\psi}  F_{1} F_{2}^2 F_{1}
    \,  \ket{\psi}} +
    \frac{\delta\tau_{2}}{2}
 \, \frac{\bra{\psi}  F_{1} F_{2}^2
    \, \si_{2} F_{1}\, \ket{\psi} }
    {\bra{\psi} F_{1} F_{2}^2 F_{1}
    \,  \ket{\psi}} + \frac{\delta\tau_{3}}{2}
 \, \frac{\bra{\psi}  F_{1} F_{2}
    \, \si_{2} F_{2} F_{1}\, \ket{\psi} }
    {\bra{\psi} F_{1} F_{2}^2 F_{1}
    \,  \ket{\psi}}
\ea

\begin{figure}[h!]
\centerline{\scalebox{.5}{\includegraphics{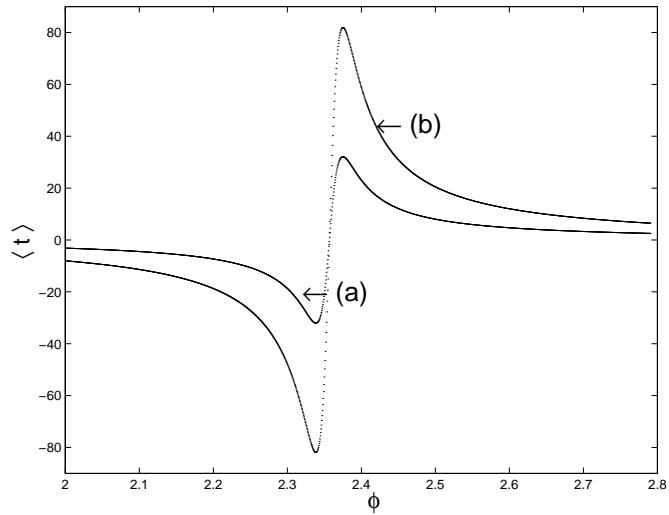}}}
\caption{Mean time-of-arrival $\langle t \rangle $ versus
polarization (we consider linear polarizations
$\ket{\psi}=\cos(\phi/2)\ket{H}+\sin(\phi/2)\ket{V}$). In this
example we used: $\delta\tau_{j}=1$, for $j=1,2,3$ and $\mu_{j}=1$
for $j=1,2$. (a) Five elements PMD-PDL-PMD-PDL-PMD network. The
PDL is distributed along the line. (b) PMD(3)-PDL(2) setup. All
PMD (PDL) sections are grouped at the beginning (at the end) of
the setup. The time shift can be made nearly three times larger
with this second setup. } \label{numres}
\end{figure}

Numerical simulations show that in order to maximize $\langle t
\rangle $, one has to align respectively all PMD's and all PDL's
and choose the PDL axis orthogonal to the PMD axis. Now one
question remains: should the PDL be distributed all along the
network or simply grouped at the end of the setup. It turns out
that the second choice is better (see Fig.\ref{numres}) and our
formula clearly shows why. From the three terms of equation
(\ref{num}) the first one is the largest, since all filters
contribute to the post-selection. But this first term is precisely
the contribution of the setup where all the PDL is at the end.
Thus $\langle t \rangle $ is maximized whenever the weight of the
first term (say $\delta\tau/2$) is the largest. This is of course
the case when all PMD's sections are put together. Since they are
all parallel, we have $\delta\tau/2= \sum_{j=1}^{N}
\delta\tau_{j}/2$.

\newpage

\section{Conclusion}

In conclusion, we have demonstrated that the formalism of weak
measurements with post-selection describes important polarization
effects in the physics of telecom optics. It is quite nice and
surprising to see that telecom engineers and quantum theorists,
two apparently completely unconnected categories of physicist,
speak of the same things, each one in his own language. We also
showed that the quantum formalism simplifies telecom calculations
and gives a better understanding of the physics of networks. It
must also be mentioned that with this work we close a loop of
analogies. On the one hand, Gisin and Go showed in \cite{gisingo}
the strong analogy between PMD-PDL effects in networks, and the
mixing and decay that are intrinsic to kaons. Remember that the
kaon system is one of the most celebrated examples of a system
evolving according to an effective non hermitian Hamiltonian. On
the other hand, it was shown in \cite{massar} that a system
coupled to another suitably pre- and post-selected system can also
evolve according to an effective non hermitian Hamiltonian. So our
work closes the loop by showing the link between PMD-PDL effects
and weak measurements with post-selection.



\end{document}